\documentclass{aa}

\usepackage{color}
\usepackage[varg]{txfonts}
\usepackage{amsmath}
\usepackage{natbib}
\usepackage{graphicx}
\bibstyle{aa}
\bibpunct{(}{)}{;}{a}{}{,}

\begin{document}

\title{Temperature gradient in the solar photosphere. Test of a new spectroscopic method and study of its feasibility for ground-based telescopes. }

  \author{
   M. Faurobert\inst{1}
   \and
  M. Carbillet\inst{1}
   \and
   L. Marquis\inst{2}
   \and
   A. Chiavassa\inst{1}
   \and
   G. Ricort\inst{1}}

   \institute{ Universit\'{e} C\^{o}te d'Azur,  Observatoire de la C\^{o}te d'Azur, CNRS UMR 7293 J.L. Lagrange Laboratory,
   Campus Valrose, 06108 Nice, France\\
              \email{marianne.faurobert@oca.eu, marcel.carbillet@oca.eu}
           \and
           ISAE-ENSMA-  Teleport 2, 1 Avenue Clement Ader, 86360 Chasseneuil-du-Poitou, France\\
           \email{lucas.marquis@etu.isae-ensma.fr}
             }
\date{}
\titlerunning{Temperature gradient in the  low solar photosphere}
 \authorrunning{Faurobert et al.}
 
\abstract
{ The contribution of quiet-Sun regions to the solar irradiance variability is currently unclear. Some solar-cycle variations of 
the quiet-Sun  
physical structure, such as the temperature gradient, might  affect the irradiance. 
The synoptic measurement of this quantity along the activity cycle would improve our understanding of long-term irradiance
variations.}
{We intend to test a method previously introduced for measuring  the photospheric temperature gradient from high-resolution spectroscopic observation 
and to study its feasibility
with ground-based instruments with and without adaptative optics.}
{We used synthetic profiles of the FeI 630.15 nm obtained from realistic three-dimensional hydrodynamical simulations of the photospheric granulation and
line radiative transfer computations under local thermodynamical equilibrium conditions. Synthetic granulation images at different levels in the line are obtained by convolution with the instrumental point spread function (PSF) under various conditions of atmospheric turbulence, with and without correction by an adaptative optics  (AO) system. 
The PSF are obtained with the PAOLA software, and the AO  performances are inspired by the system that will be operating on the Daniel
K. Inouye Solar Telescope.
}
{ We consider two different conditions of atmospheric turbulence, with  Fried parameters of 7 cm and 5 cm, respectively. 
We show that the degraded images lead to both a bias and a loss of precision in the temperature-gradient measurement, and that  the correction with the AO system allows us to drastically improve
the measurement quality.}
{ Long-term synoptic observations of the temperature gradient in the solar photosphere can be undertaken by implementing this method on ground-based solar telescopes that are equipped with an AO correction system.
}
 

\maketitle

\section{Introduction}

The variation of the total solar irradiance (TSI)  in phase with the solar cycle is now well established thanks to measurements
performed in space by dedicated
instruments over the past four solar cycles. However, the long-term variations of the TSI at solar minimum are still under debate because different groups that use different composite data
 find either a constant or a varying TSI at solar minimum. 
Moreover, the spectral dependance (SSI) of the solar cycle variability is also a 
controversial subject  \citep[see the reviews by][]{Yeo2014, solankiIR2013}. The direct observation of SSI variations on solar-cycle timescales is a difficult task, and some discrepancies appear, in particular in the UV domain, between different space-based instruments.  
These variations are  an important forcing term for  the evolution of the climate
on Earth.
The  efforts made to model TSI and SSI variations currently rely  on the hypothesis that the main drivers of solar variability are magnetic features at the 
solar surface, and that the thermodynamical state of the quiet-Sun atmosphere is invariant. Irradiance reconstructions based on the effect of magnetic concentrations 
at the solar surface allow recovering more than 90\% of the observed variations on this timescale. Very recently, \citet{bib121} made a step forward in TSI modeling by using realistic three-dimensional magnetohydrodynamics (MHD) simulations of the solar magnetic concentrations instead of one-dimensional semi-empirical models. The model replicates 95\% of the observed variability between April 2010 and July 2016 over the examined timescales (days to years). 

Other plausible mechanisms implying a modulation of the global structure of the Sun have also been proposed. 
 Ideas on global structural changes are supported by the observation of the frequencies of the acoustic p-modes
that also show a well-documented solar-cycle variation  \citep[see ][]{fossat1987, salabert2015}. Recent  
global MHD simulations of the solar convection zone are presented in \citet{bib107}.  Cyclic large-scale axisymmetric magnetic fields with polarity reversals on decadal timescales are recovered by the simulations, and they show that the thermal convective luminosity varies in phase with the magnetic cycle. However, the simulation 
domain stops below the photosphere, which limits the type of comparisons that could be made with the real Sun.

Theoretical investigations  based on MHD simulations
 of the properties of the granulation in quiet regions of the Sun in different magnetic environments   were
presented in 
 \citet{criscuoli2013}.  They showed that an ambient magnetic field gives rise to two main physical effects:  it inhibits convection, and it reduces the average opacity. 
 The two effects have opposite consequences on the temperature gradients: the reduced opacity, which dominates at optical depths close to unity and smaller,
reduces the temperature gradient; but the inhibited convection, which dominates at optical depths larger than one, steepens the temperature gradient.  
In internetwork regions, the simulations showed that the  temperature in the low photosphere decreases with increasing  local magnetic field 
\citep[see also][]{criscuoliUitenbroek2014}. 
This was later confirmed by \citet{Faurobert2016} (paper I)  where a spectroscopic method for measuring the temperature gradient in the low photosphere was 
presented and applied to Hinode/SP observations in the FeI 630.15 nm line. The analysis of two sets of observations obtained in December 2007 and December 2013, respectively,
showed a decrease in temperature in the low photosphere at the solar maximum in 2013.
This finding must be confirmed on a larger set of observations made throughout the solar cycle. 

Possible contributions of the quiet Sun to long-term variations of the irradiance 
could be investigated by a systematic follow-up of the photospheric temperature gradient. 
As it seems very difficult to ensure constant and controlled instrumental performances over more than one solar cycle on a space-based instrument, 
synoptic programs dedicated to the measurement of the photospheric temperature gradient should rather be undertaken on ground-based instruments. 

Here we  present some tests of the measurement method presented in Paper I and a  feasibility study for ground-based observations under different conditions of
turbulence in the Earth atmosphere, with and without corrections by an AO system. The image degradation at a telescope focus due to  atmospheric turbulence is 
related to the so-called Fried-parameter $r_0$ , representing the coherence scale of the turbulent motions. Here we used two values for $r_0$,  $r_0$= 7 cm and $r_0$=  5 cm.
 7 cm is the median value measured at the Haleakala site where the Daniel K. Inouye Solar Telescope (DKIST) is currently under construction, and the case with $r_0$ = 5 cm allows us to investigate 
 more severe turbulence conditions. The point spread functions (PSFs)  in the different situations are obtained from the PAOLA sofware 
 \citep{Jolissaint2006, Jolissaint2010}, and for the characteristics of the adaptative optics (AO), we refer to the description of the system that will operate on DKIST, which was presented 
 in \citet{DKISTAO12016} and \citet{DKISTAO22016}
 
 In Section 2 we explain how the synthetic images of the granulation at 
 different line levels are obtained under different atmospheric turbulence conditions, and we briefly recall the temperature-gradient measurement method (a more detailed presentation 
 is given in Paper I). Section 3 is devoted to the presentation of the results.

\section{Measurement method and synthetic images}

\subsection{Synthetic images}

\begin{figure}[ht]
\includegraphics[width=0.45\textwidth]{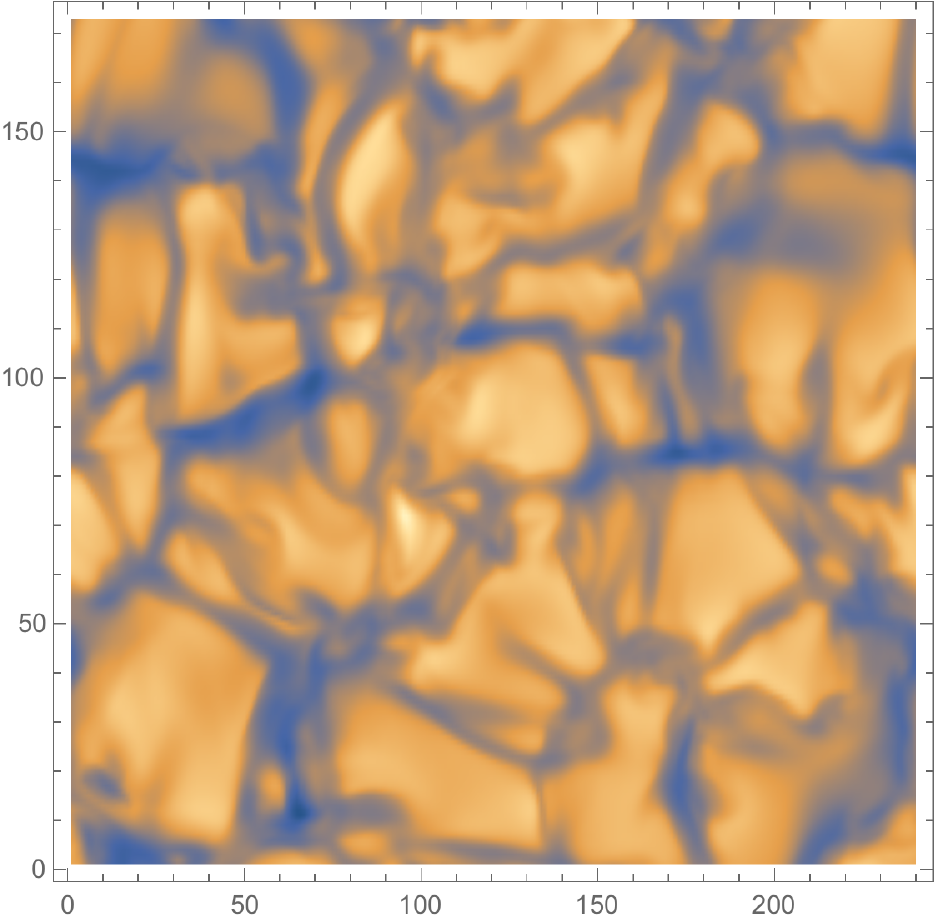}

 \caption{Synthetic image of the granulation in the 630 nm continuum at $\cos\theta =0.85$. The pixel size is 0.045''.}
 
  \label{fig1}
\end{figure}

We used a radiative hydrodynamical (RHD) simulation of the Sun taken from the stagger grid \citep{magic2013}. It was computed using the stagger code, a
dedicated (magneto)hydrodynamics code that solves the time-dependent equations
for conservation of mass, momentum, and energy. The simulation has 240x240x240 grid points corresponding to a horizontal extension of 8 Mm and a horizontal resolution of 33.3 km.
Then we used the pure local thermal equilibrium (LTE) radiative transfer code \textsc{Optim3D}  \citep{chiavassa2009} to compute synthetic spectra from the snapshots of the RHD  simulation. Zero microturbulence was assumed, and the nonthermal Doppler broadening of spectral lines only uses the self-consistent velocity fields  directly from the three-dimensional simulations. The temperature and density ranges spanned by the tables are optimized for the values encountered in the RHD simulation. The detailed methods used in the code are explained in \cite{chiavassa2009}. 
The spectra were computed sampling 115 wavelengths in the interval [630.08 nm , 630.32 nm] with a sampling of 2.1 pm and for different inclinations with respect to the vertical. 
For a given inclination angle (location on the solar disk), we obtain cubes of simulated $(x,y, \lambda )$ data that represent  solar scenes. 
We show in Fig. \ref{fig1} a continuum image obtained at the heliocentric angle $\theta$ with $\cos\theta =0.85$ for one snapshot of the RHD simulation.

\subsubsection{Without atmospheric turbulence}

\begin{figure}[ht]
\includegraphics[width=0.45\textwidth]{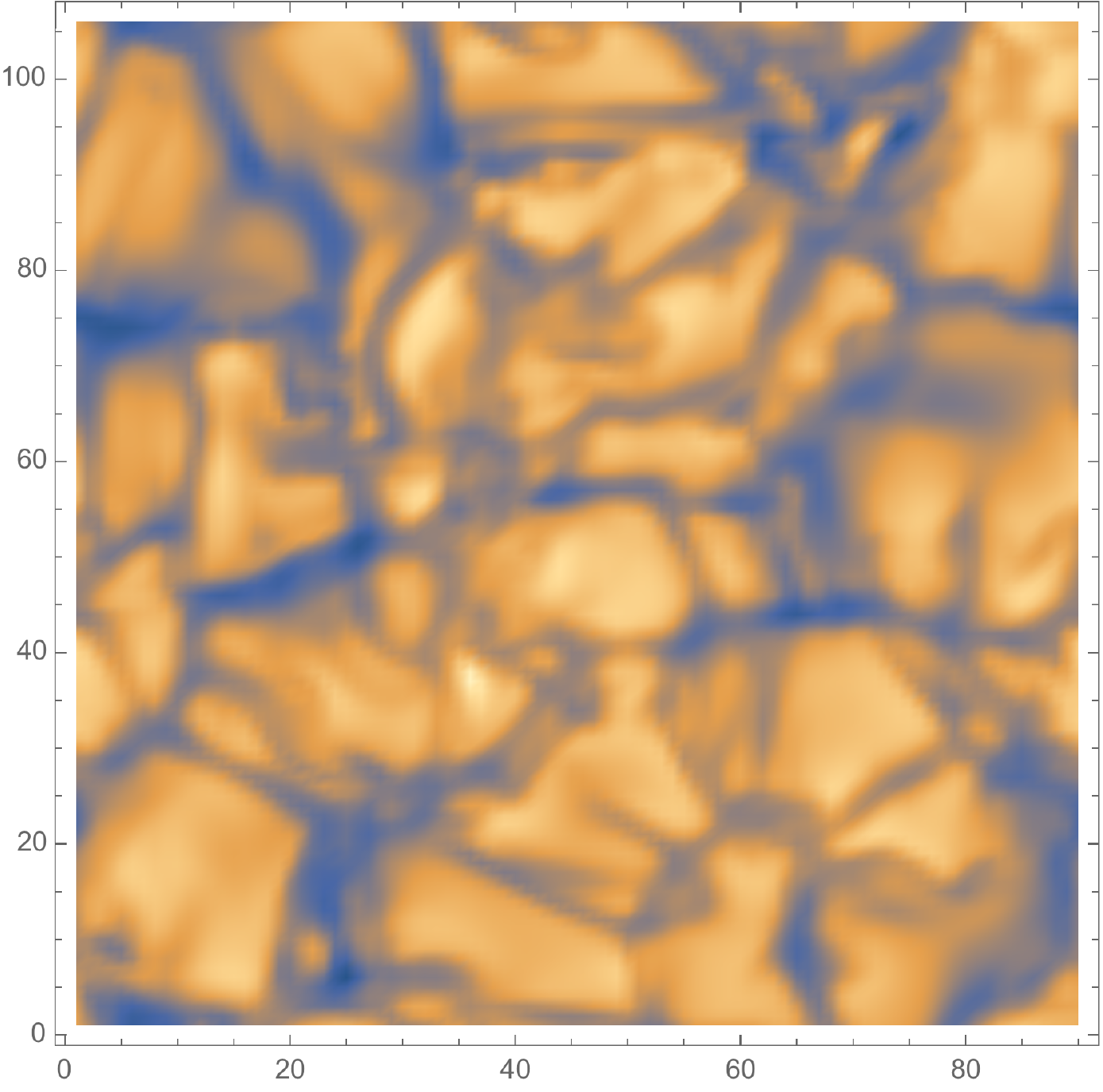}

 \caption{Same image as in Fig. \ref{fig1} after diffraction by the telescope pupil and rebinning to a pixel size of  0.106''. }
 
  \label{fig2}
\end{figure}

To obtain synthetic cubes of  $(x,y, \lambda )$ data observed with a telescope, we need to convolve the solar scene with the PSF of the instrument.  In a first step, we considered only the image degradation due to the diffraction by the telescope pupil and by the spectrograph slit. For the diffraction, we assumed that the telescope has a 4m diameter off-axis mirror like the DKIST, and the spectrograph profile is a Gaussian function with a half-width of 2.1 pm. Under these very favorable conditions, the spatial resolution $\lambda/D = 0. 0325'' $) 
is slightly better than the resolution of the numerical simulation, and the line profiles are only slightly broadened. To compute the convolution of the images with the Airy function, we oversampled the simulated images by a linear interpolation between the grid points. Then we rebinned the images to a pixel size of 0.106'' to account for the width of the spectrograph slit that would be used for this observing program. We also added photon noise and readout noise to the simulated
images. To estimate the mean number of photons on each pixel, we assumed an exposure time of 0.5 s,  a 4m telescope, and a pixel size of 0.106''. For the readout noise, we took  a value of 20 electrons.  Photon noise and readout noise are negligible for such an exposure time. 
The resulting image in the continuum is shown in Fig. \ref{fig2}.

\subsubsection{Degraded and AO-corrected images}

The previous ideal case is not realistic as turbulence in the Earth atmosphere always degrades the images. To model the PSF of the system (atmosphere+telescope+AO system), we made use of the analytical code {\tt PAOLA} \citep{Jolissaint2006, Jolissaint2010}, which adopts a synthetic approach for modeling the performance of an astronomical AO system: the global behavior of the whole system at once is modeled with the help of an analytic expression for the residual phase average (or long-exposure) spatial power spectrum and its relationship with the long-exposure AO optical transfer function. The main advantage of this approach, with respect to a detailed end-to-end Monte Carlo-based approach, is the gain in computation time, which can be particularly enormous in the present case of wide-field correction. We note that an inter-validation study between {\tt PAOLA} and the well-known end-to-end Monte Carlo-based software tool {\tt CAOS} \citep{Carbillet2005} shows a very good agreement for the fundamental fitting error and anisoplanatic error \citep{CarbilletJolissain}, which are expected to dominate in the present wide-field AO case.

\begin{figure}[ht]
\includegraphics[width=0.9\columnwidth]{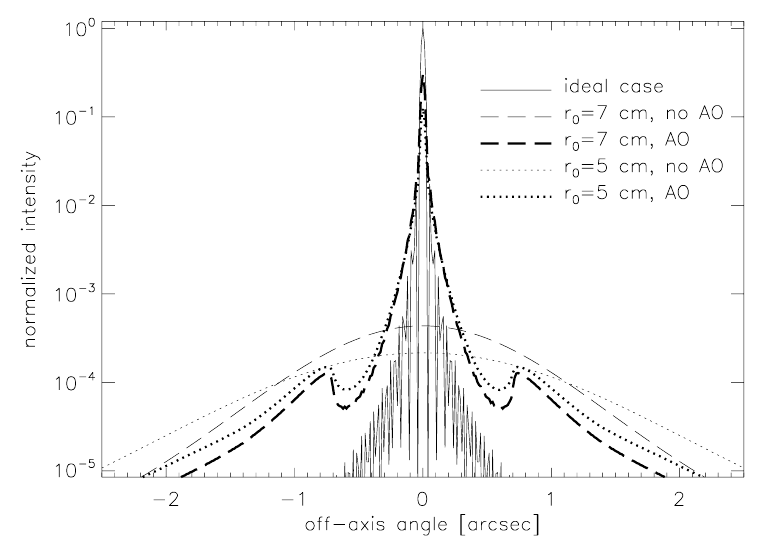}
 \caption{Logarithmic plot of the simulated post-AO and seeing-limited PSFs for the two cases considered ($r_0$=7\,cm and $r_0$=5\,cm) and compared with the ideal PSF.
 }
  \label{figPSF}
\end{figure}

\begin{figure}[ht]
\includegraphics[width=0.45\textwidth]{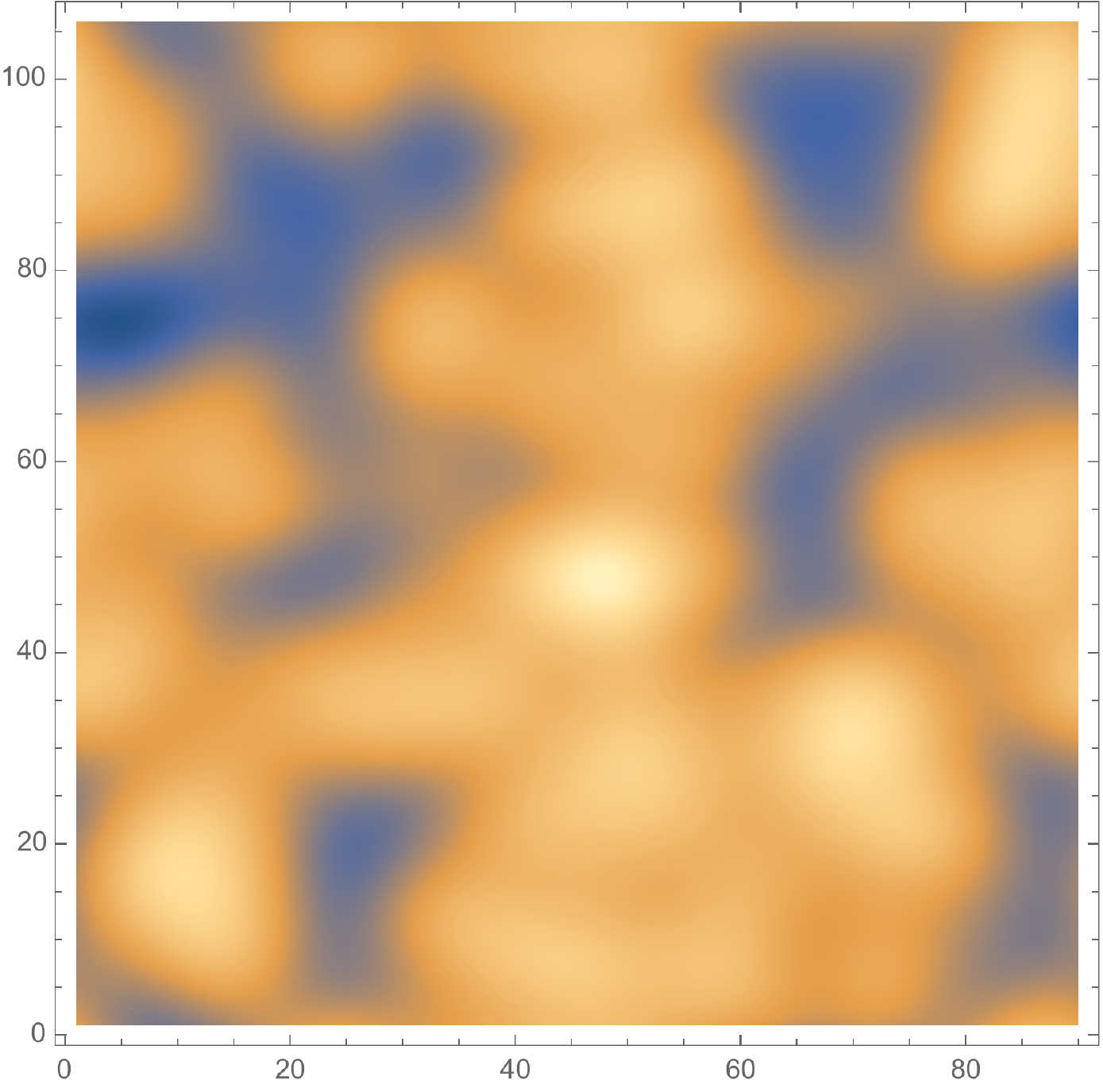}
\includegraphics[width=0.45\textwidth]{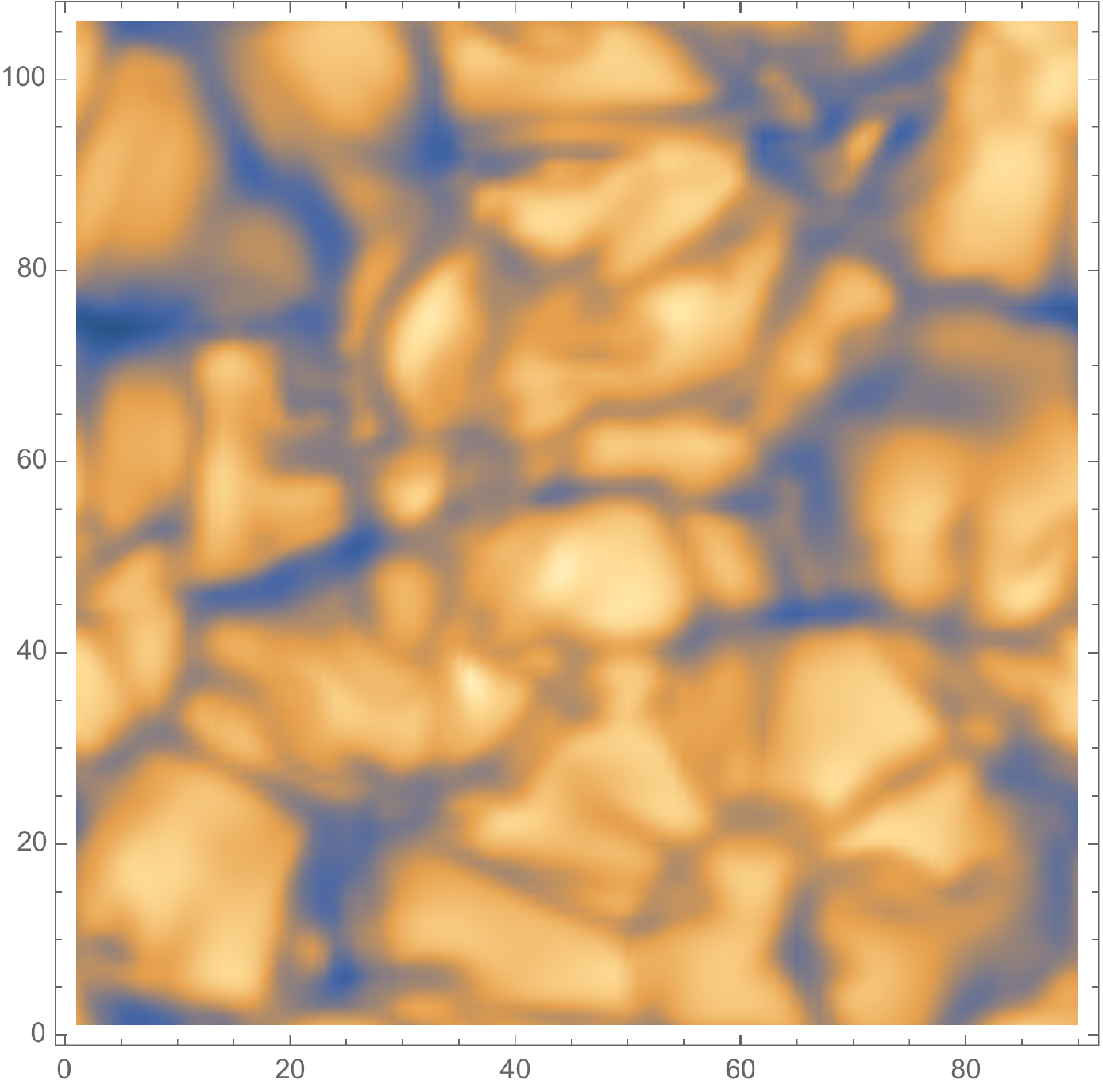}
 \caption{Upper panel: Continuum image obtained without AO for atmospheric turbulence  $r_0 = 7$ cm for one snapshot of the granulation. Lower panel:
 Continuum image with AO for the same $r_0$ value. The pixel size is 0.106''.
 }
  \label{figOA1}
\end{figure}

\begin{figure}[ht]
\includegraphics[width=0.45\textwidth]{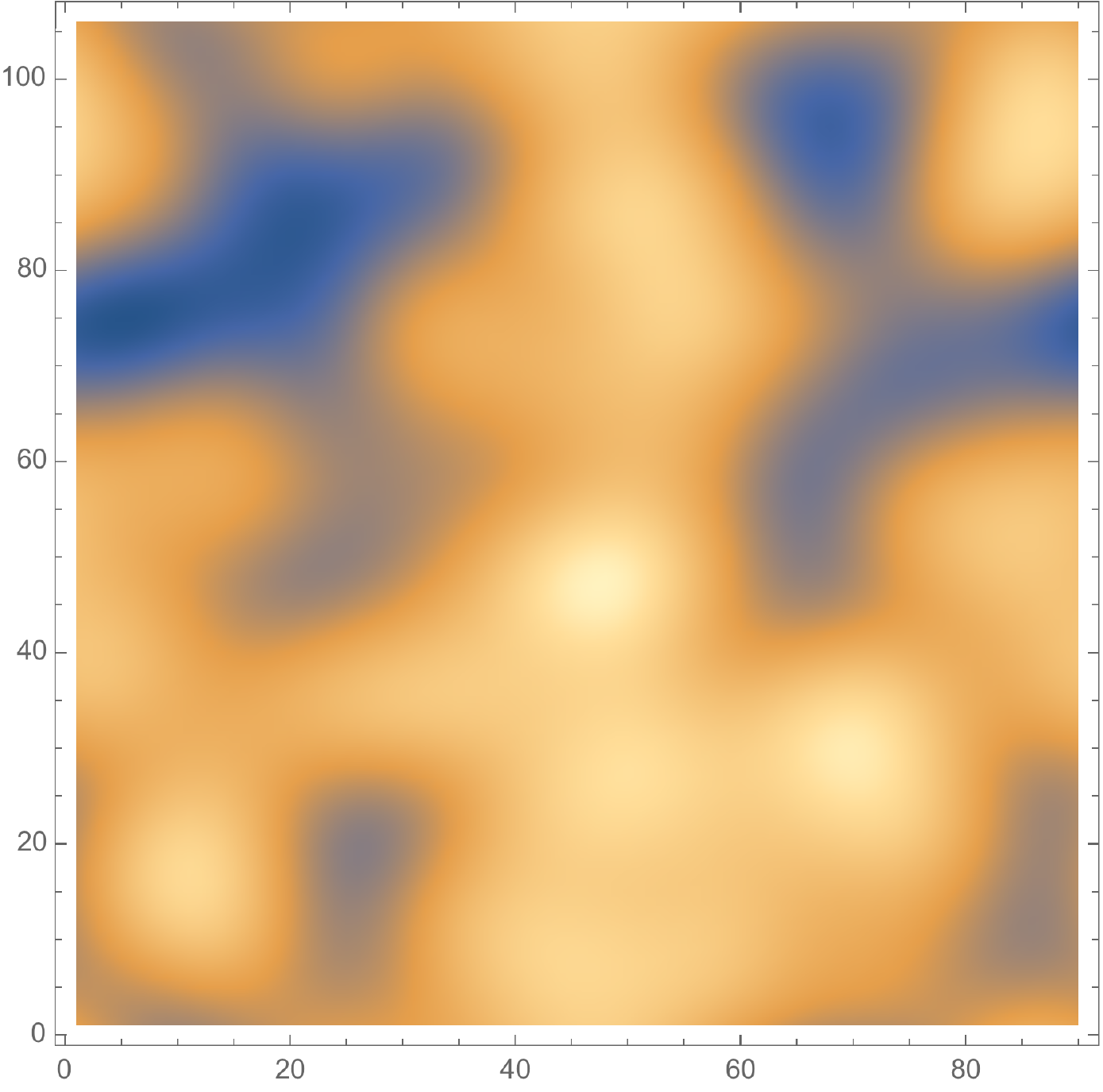}
\includegraphics[width=0.45\textwidth]{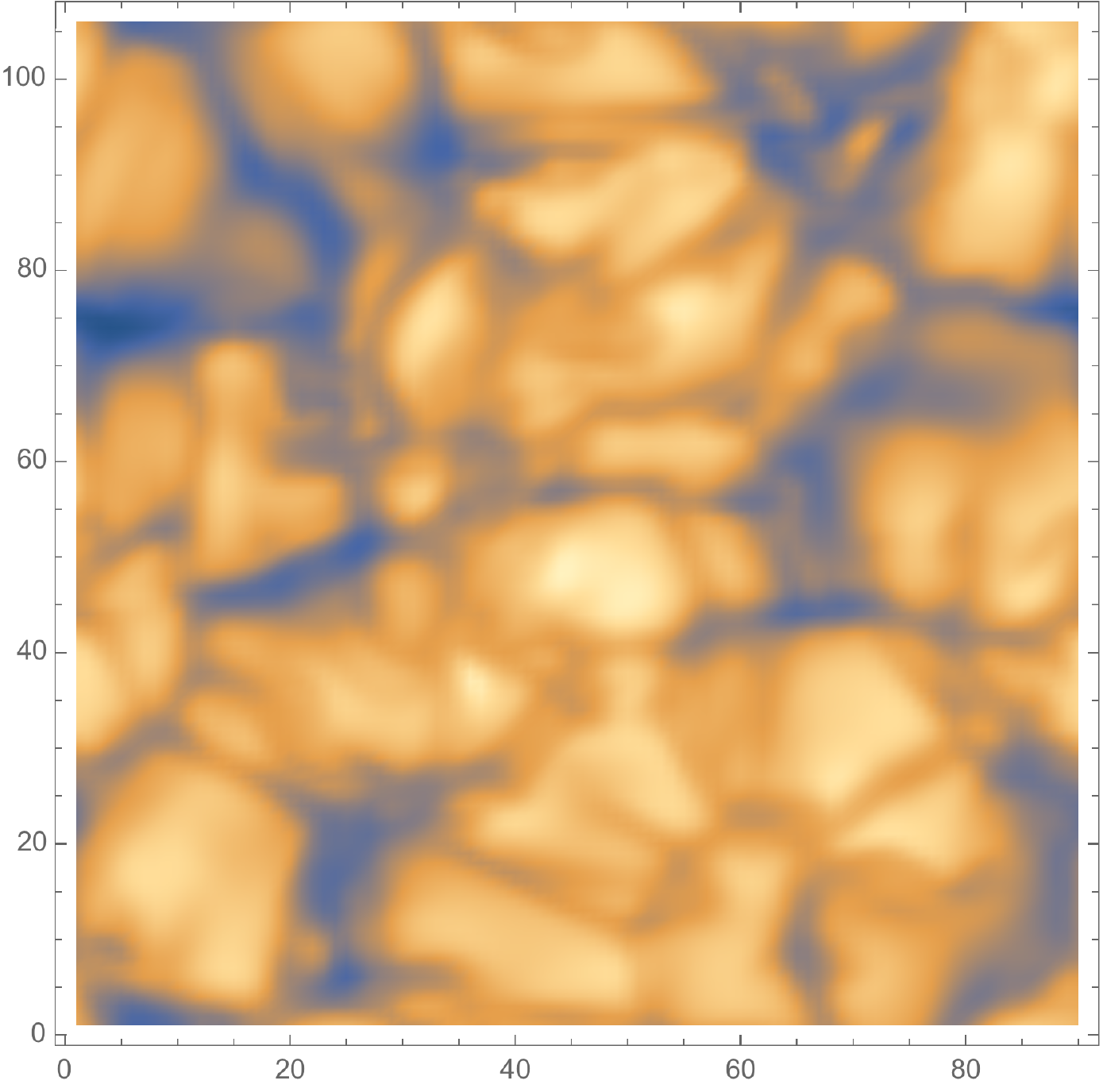}
 \caption{Same as Fig. \ref{figOA1} for  $r_0 = 5$  cm.
 }
  \label{figOA2}
\end{figure}

The (relevant) case study of the DKIST and its wide-field AO system being developed has been considered here, and PAOLA was used in its GLAO "full" mode, in which the phase is assumed known in the full sensing field of view defined here by a circle of diameter 10\arcsec. Its atmospheric and AO system parameters were taken from \cite{DKISTAO12016} and \cite{DKISTAO22016}, with an expected Fried parameter $r_0=7$\,cm and a related on-axis Strehl ratio goal of 0.3 at 500\,nm. The exact composition of the three-layer median profile taken into account by these authors was kindly communicated to us by Jose Marino. With respect to these assumptions, we have chosen to add a quantity of anisoplanatic error throughout the field of observation, leading to a Strehl ratio of 0.3 as well, but at 630\,nm, and we assumed that the resulting PSF is stable throughout the field of observation. In addition, we considered a worse-case scenario with $r_0=5$\,cm and the same AO system parameters, leading to a Strehl ratio of 0.12 at 630\,nm. The parameters of the resulting post-AO PSF modeling are reported in Table\,\ref{tab:AO}. These two long-exposure post-AO PSFs, together with the two corresponding seeing-limited PSFs and the ideal PSF (no perturbation at all), are presented in Fig.\,\ref{figPSF}. They were then used for convolution with the various previously modeled solar scenes in order to obtain the final observed images (assuming stability of the correction on the whole field of observation). The resulting images in the continuum are shown in Fig.\,\ref{figOA1} ($r_0$=7\,cm) and Fig.\,\ref{figOA2} ($r_0$=5\,cm).

\begin{table}
\begin{center}
\caption{Main parameters of the post-AO PSF modeling.}
\vskip .25cm
\begin{tabular}{lr}
\hline
{\bf turbulent atmosphere parameters}                   &                                                               \\
Fried parameter $r_0$ (at 500\,nm)                              &   7\,cm and 5\,cm                                       \\
$\Rightarrow$ resulting seeing angle (at 500\,nm)       &   1.44\arcsec and 2.02\arcsec         \\
number of turbulent layers                                              &   3                                                             \\
layer altitudes                                                                &   [0, 7852, 13252] m                            \\
layer velocities                                                       &   [10, 10, 10] m/s                                      \\
layer $C_N^2$ profile relative weight                  &   [0.942, 0.038, 0.02]                            \\
wavefront outer scale ${\cal L}_0$                             &   27\,m                                                 \\ \hline
{\bf telescope parameters}                                              &                                                               \\
effective diameter                                                      &   4\,m                                                  \\
obstruction ratio                                                       &   none (off-axis mirror)                                \\ \hline
{\bf AO system parameters}                                      &                                                               \\
deformable mirror configuration                                 &   44$\times$44                                  \\
Shack-Hartmann configuration                                    &   43$\times$43                                  \\
wavefront sensing exposure time [ms]                    &   2                                                           \\
additional AO-loop delay time [ms]                              &   0.5                                                   \\
wavefront sensing RON [e$^-$\,$rms$]                    &   20                                                  \\
additional anisoplanatic off-axis angle                         &   2.6\arcsec                                            \\
closed-loop gain                                                        &   0.5                                                   \\
$\Rightarrow$ resulting Strehl ratio (at 630\,nm)       &   0.30 and 0.12                                        \\ \hline
\label{tab:AO}
\end{tabular}
\end{center}
\end{table}
  
\subsection{From spectrograms to images at constant continuum optical depth }

In a spectral line, the  opacity varies steeply with the  wavelength.  It is 
 given by
 \begin{equation}
 k (\lambda)= k_c +k_l \phi( (\lambda-\lambda_0)/\Delta\lambda_D),
 \end{equation}
 where $k_c$ is the continuum opacity,  $k_l$ is the frequency-averaged line opacity, $\lambda_0$ is the line center wavelength in the observer frame, and $\phi$ is the normalized absorption profile, 
 which in the general case is given by the Voigt function, and   $\Delta\lambda_D$ is the line Doppler width. In the following we use the notations $\delta\lambda= (\lambda-\lambda_0)/\Delta\lambda_D$ and $r=k_l/k_c$.
 The Doppler broadening is due both to thermal broadening and to unresolved velocities. In the far wings, radiative and collisional  damping become dominant. 
  
The   absorption profile varies over the solar surface because of temperature, velocity, density,  and magnetic inhomogeneities, and  the line central wavelength  varies as a results of
 the varying 
global convective motions of the matter with respect to the observer. The line central wavelength is defined as the position of the minimum of the line-intensity profile. 

We first  remark that  at small line depressions,
the shape of the intensity profile is given by $I(\delta\lambda)\simeq I_c(1-r\phi(\delta\lambda ))$, where $I_c$ denotes the continuum intensity. It is therefore related to the local
values of $r $ and of the line-absorption profile. When we consider for example the image obtained at 2\% line depression, it is formed at a constant continuum optical depth surface with 
$r\phi(\delta\lambda )=0.02$. 
The line-cord width at 2\% line depression is denoted by  $\Delta\lambda_c$, and its value depends on  $r$ and the line absorption profile.

\begin{figure*}[ht]
\includegraphics[width=0.32\textwidth]{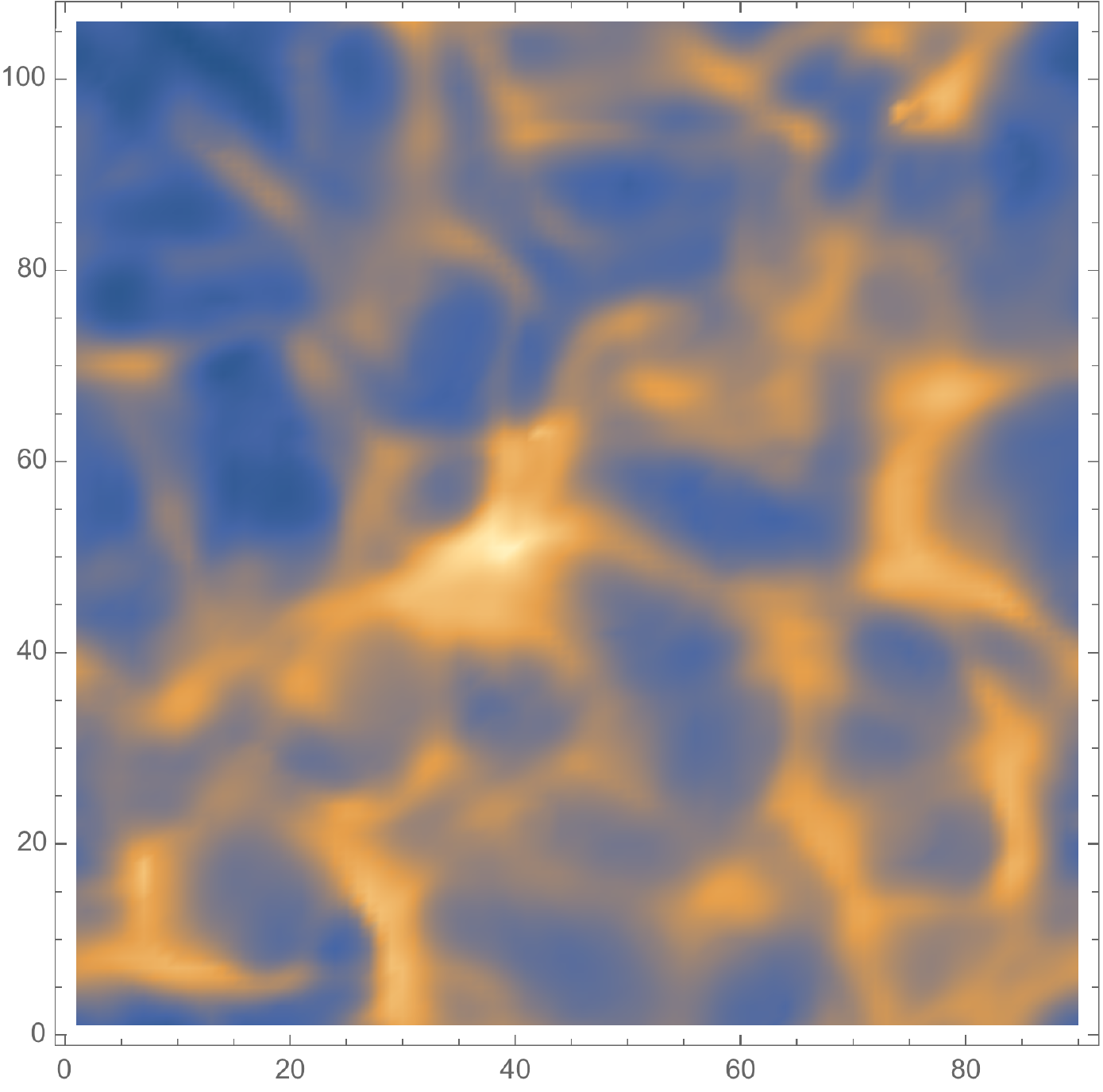}
\includegraphics[width=0.32\textwidth]{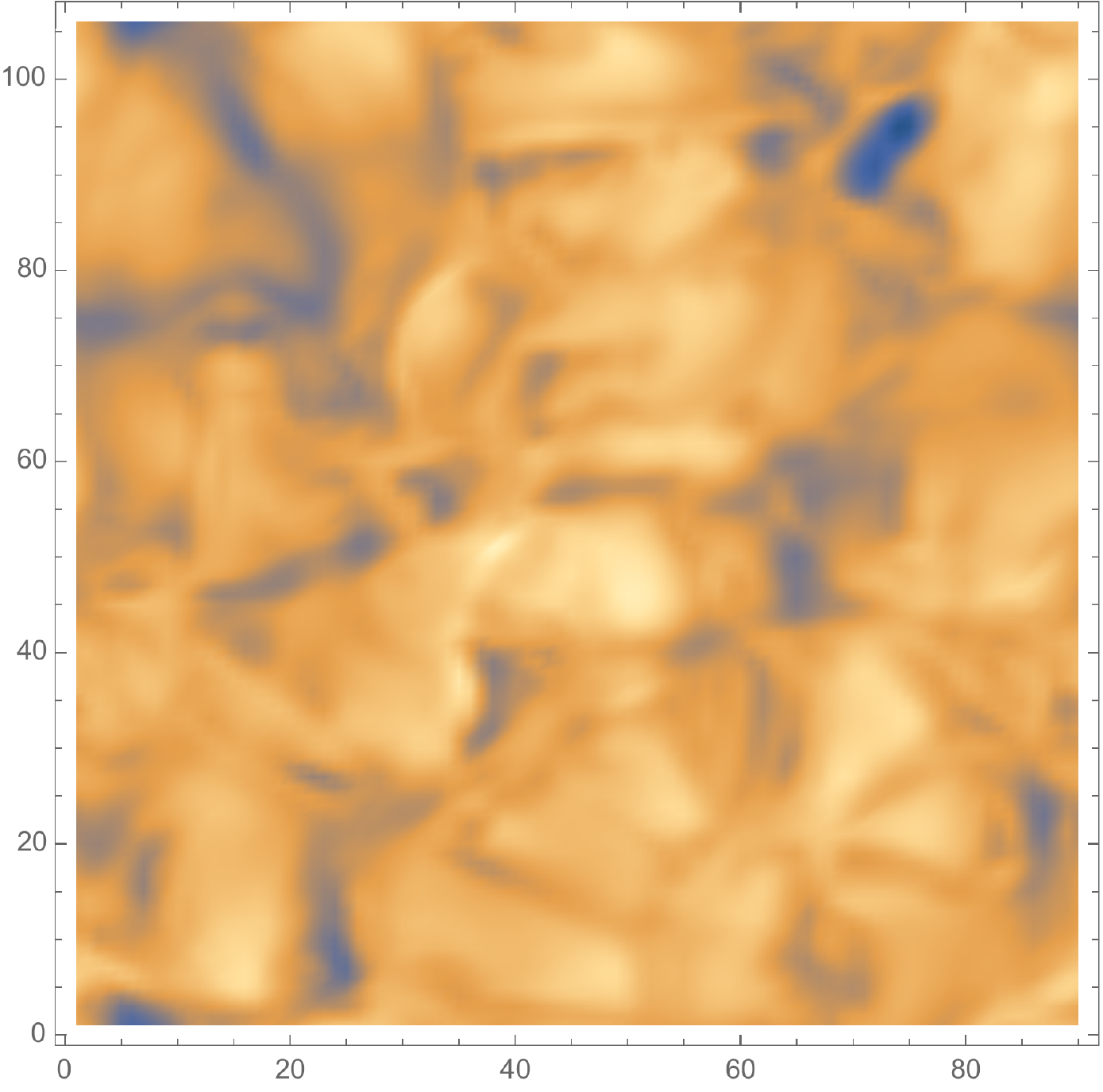}
\includegraphics[width=0.32\textwidth]{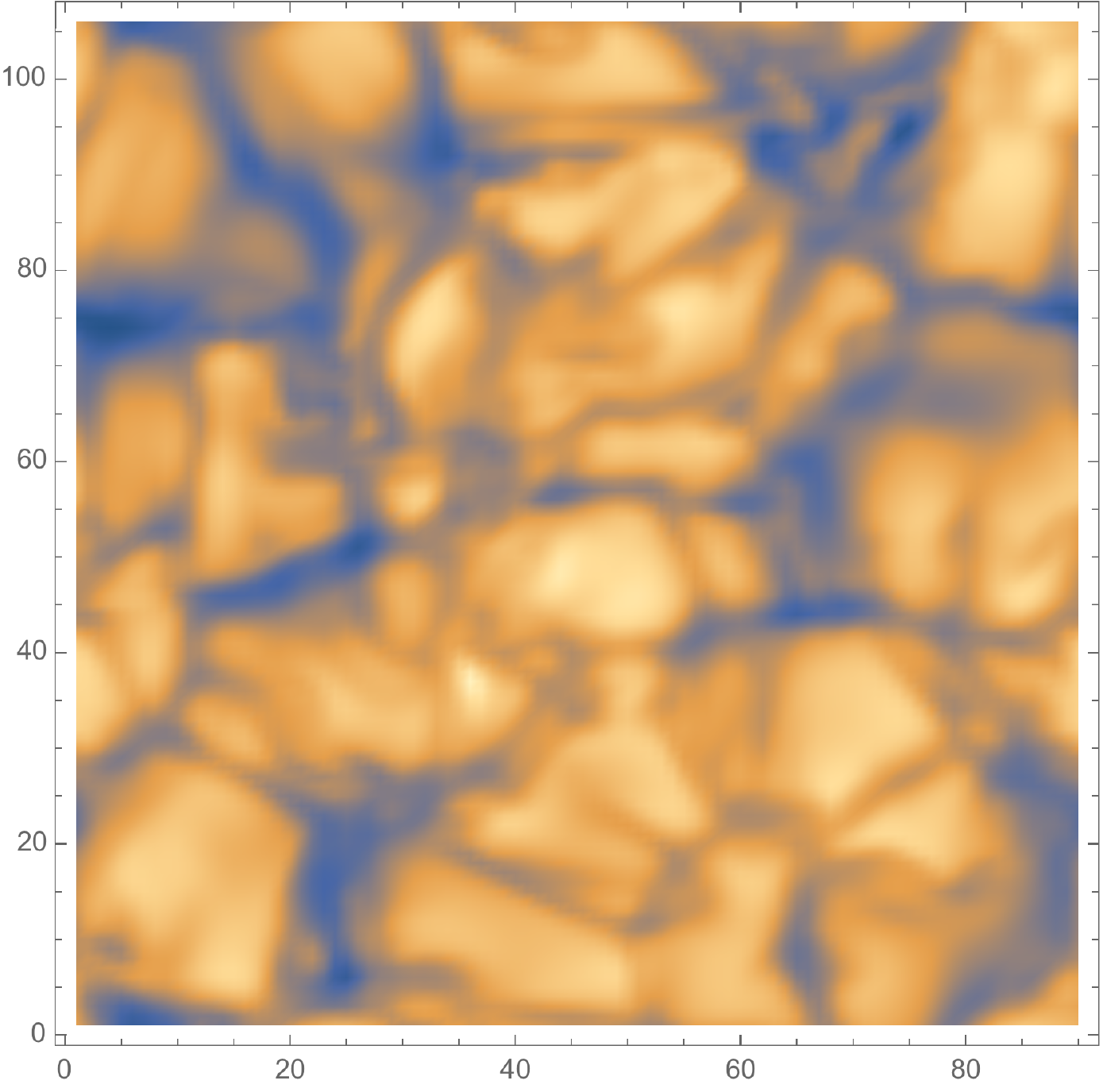}
 \caption{Images  at different line levels. The image degradation due to the atmospheric turbulence $r_0 =$ 7 cm 
 is corrected by the AO system described in the text.
  Left panel: Line level 1 (line core). Middle panel: Line level 8 (granulation-contrast inversion zone). Right panel: Line level 15 close to the continuum.
 }
  \label{fig3}
\end{figure*}

In the damping wings of the line, the profile function decreases like $1/\delta\lambda^2$, so that at $\delta\lambda=\alpha \Delta\lambda_c$, with $\alpha <1$,
we have $r \phi(\alpha \Delta\lambda_c )= (1/\alpha^2) r\phi(\Delta\lambda_c )=0.02/\alpha^2$. 
The ratio of line to continuous absorption is thus constant over an image at constant $\alpha$. 
We may therefore consider that a line-wing image obtained at constant $\alpha$ is formed on a surface where the continuum optical depth is a constant. 
This approach is valid when  $r$ and the line-absorption profile
 are depth independent or do not vary significantly 
 within the line-wing formation height. For strong lines such as the FeI 630.15 nm line,  
 this would not apply in the line core, which is formed in the upper photosphere.
 In the following we use this image reconstruction method only in the line-damping wings.
 To obtain a fine enough depth grid, we chose to define 25 line cords  given by $\delta\lambda_i=(i-1)\Delta\lambda_c/24 $. 
 The damping-wing images are typically at levels 15 to 25, nd they are well correlated with the continuum image.
 Figure \ref{fig3} shows the images obtained at three different line-levels in the case where the Fried parameter is $r_0$= 7cm, and with AO-correction.  
 
 \subsection{Temperature measurement}
 
\begin{figure}[ht]
\includegraphics[width=0.45\textwidth]{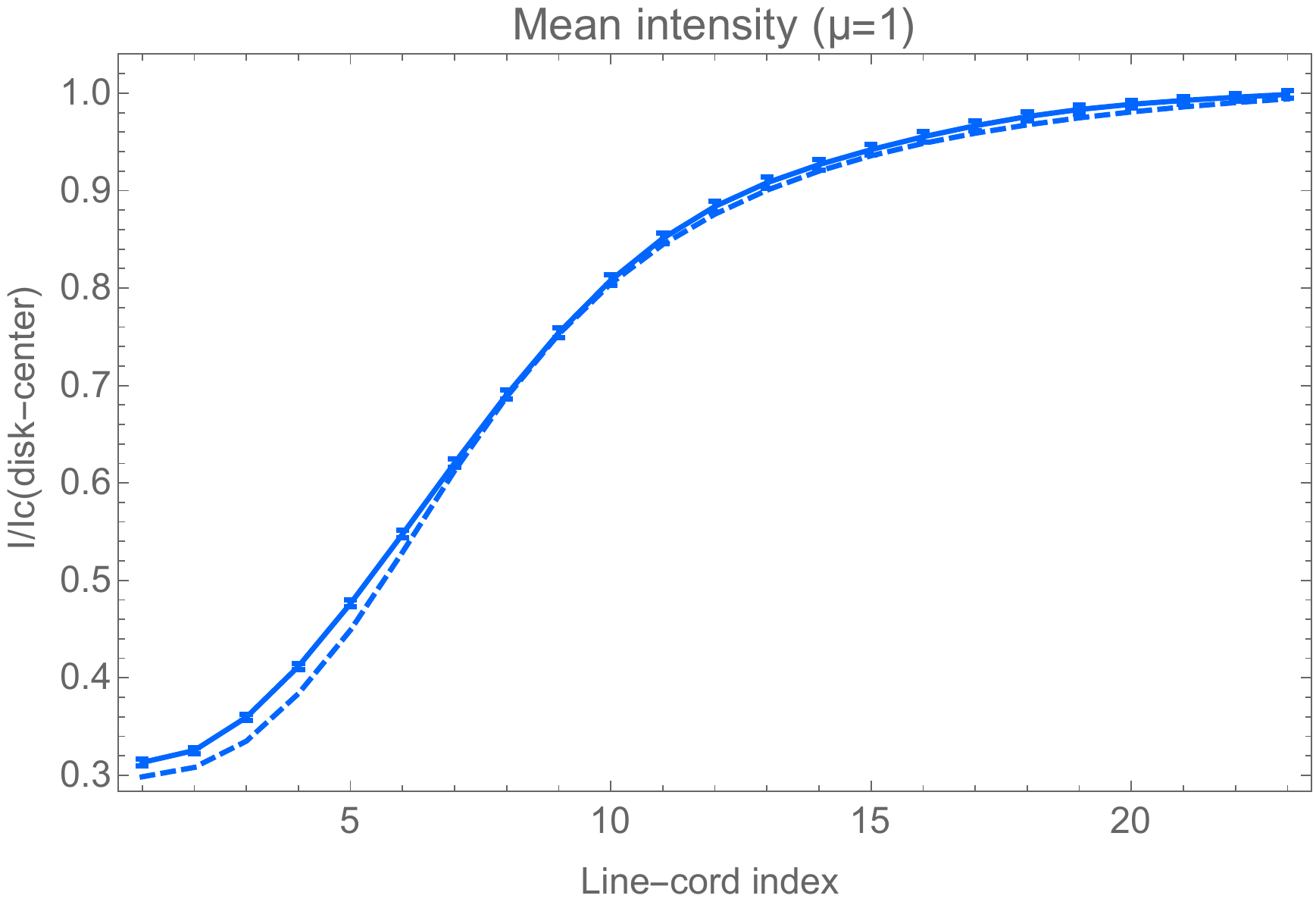}
\includegraphics[width=0.45\textwidth]{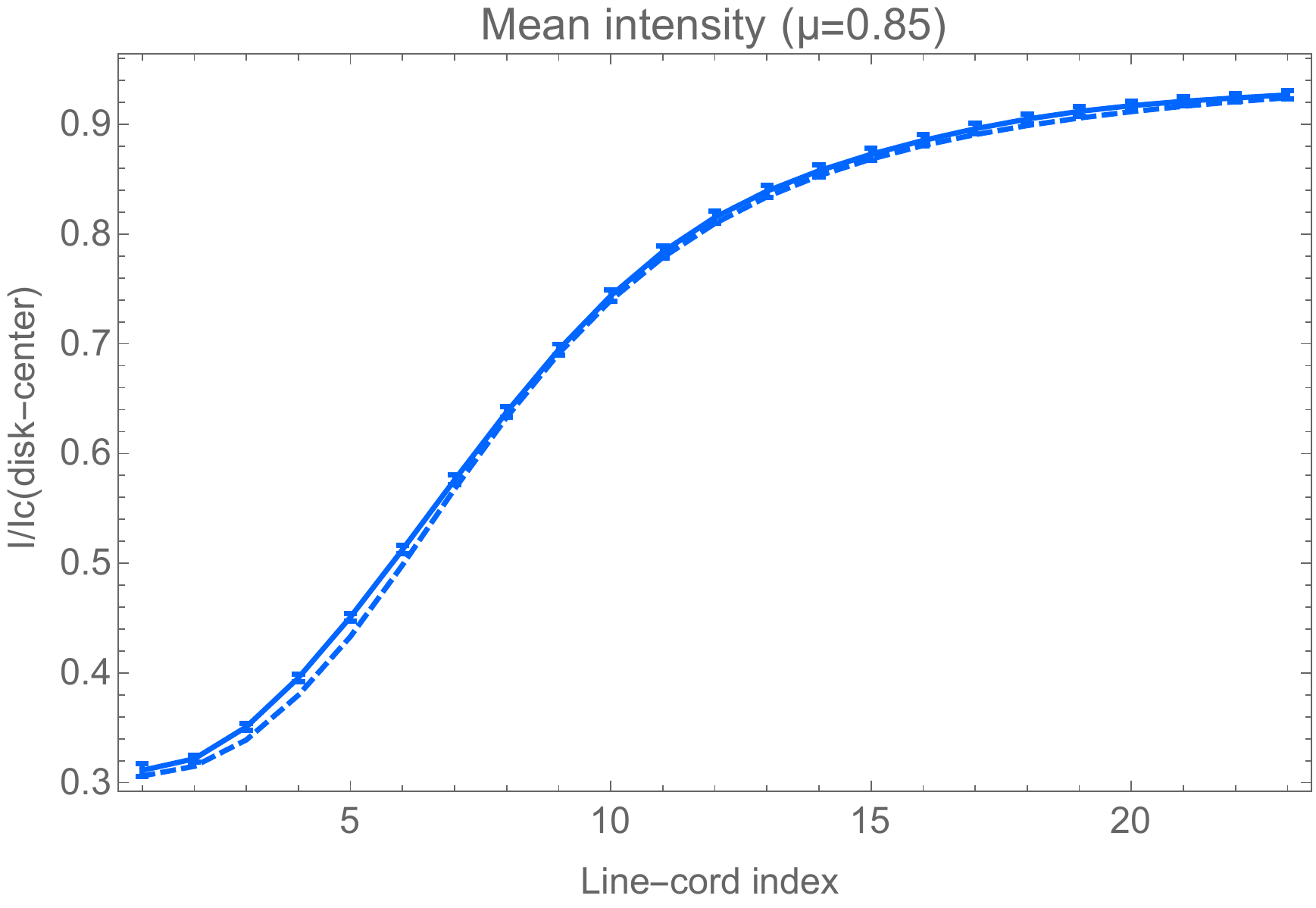}
 \caption{Intensity averaged on a 10'' x 10'' region of the quiet Sun at the 25 line levels  normalized to the average continuum intensity at disk center. 
 Dashed line: Simulations. Full lines: Hinode data.
 Upper panel: At solar disk-center. Lower panel: At $\cos\theta =0.85$
 }
  \label{fig4}
\end{figure}

 We wish to measure the mean temperature on constant continuum optical-depth surfaces. Assuming  that the FeI 630.15 nm is formed under LTE conditions in the solar photosphere, we may derive the mean temperature from the mean intensity in the images at the different line-levels. The LTE assumption has been tested by various authors \citep[see ][]{Shchukina2001} who showed that it gives very good results for the average line-profile of the FeI 630 nm line pair on surfaces of several arcseconds$^2$. 
 We also tested the LTE assumption for the average intensity obtained at the different line-levels in the FeI 630.15 nm by comparison with Hinode SOT/SP observations. Figure \ref{fig4} shows this comparison for observations of the quiet Sun at the solar minimum in 2007. 
 For Hinode data the averages
 are obtained on 10'' x 10'' surfaces away from network elements. The small differences observed at disk center close to the line center, between the LTE simulations and the observations, might be due to
 some scattered light in the SOT/SP spectrograph. The agreement is very good at line-levels larger than 10, however, where scattered light (if any) is negligible.
 
 In the following we use synthetic images at line-index larger than 14 at $\cos\theta =0.85$. 
 The mean temperature is derived from the mean intensity in the image using the Planck law. 
 The calibration of the intensity in the simulation to physical units is made using the intensity measured by \citet{Neckel2005} in the solar continuum at 630 nm
 at $\cos\theta =0.85$: $I_c(\lambda = 630 $nm, and $\cos\theta =0.85)$ = 2.76675$ \,10 ^{14}$ in cgs units.
 
  \begin{figure}[ht]
\includegraphics[width=0.45\textwidth]{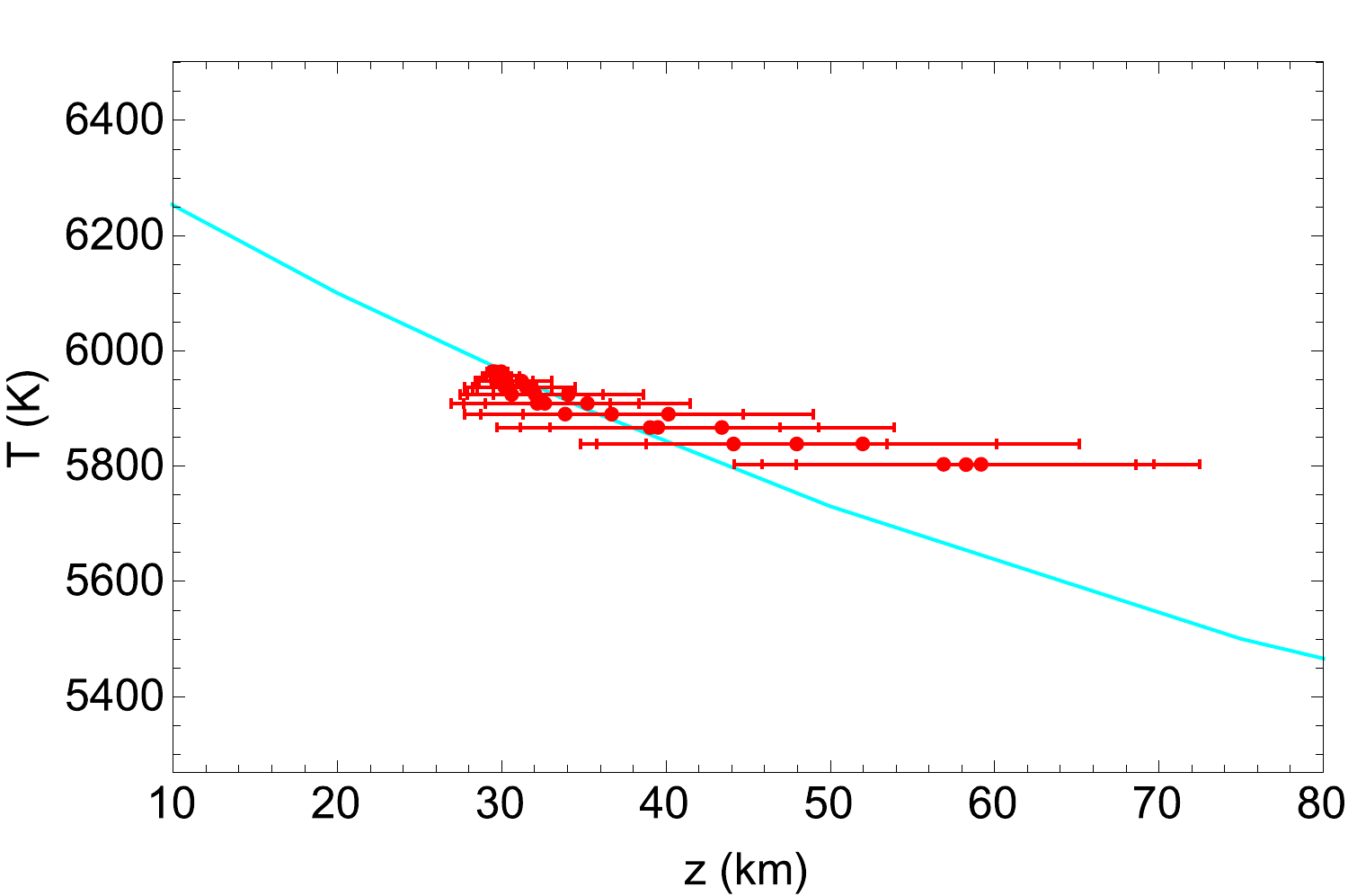}
\includegraphics[width=0.45\textwidth]{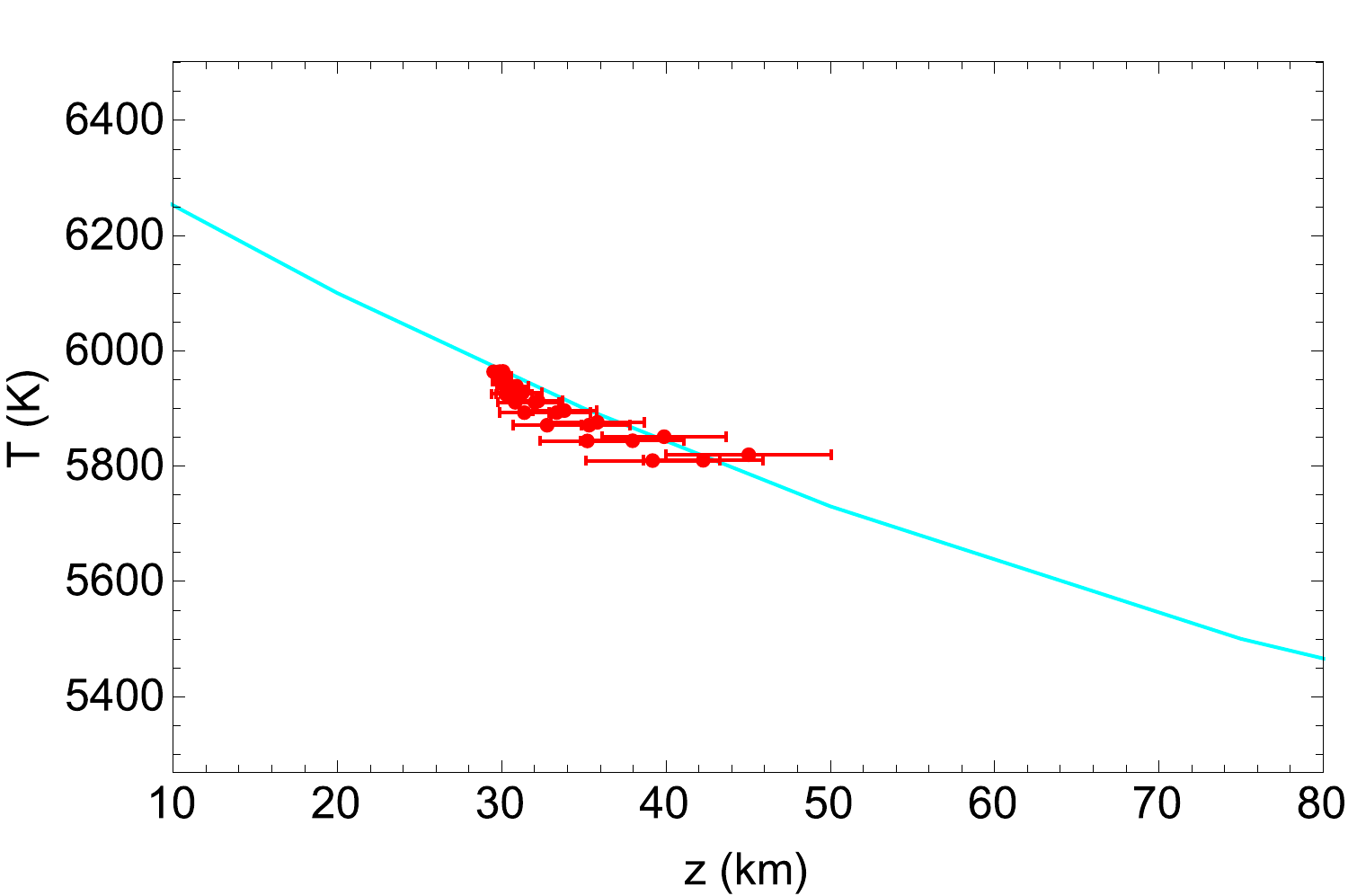}
 \caption{ Temperature as a function of altitude  for three snapshots of the RH simulation. The bars
 show one standard deviation on the measurement of the altitude. The blue curve shows the model 101 of \citet{fontenla2011}.
 Upper panel: Measurements from perturbed images with $r_0$= 7 cm. Lower panel: Measurements from images corrected with the AO system.
 }
  \label{fig5}
\end{figure} 

\begin{figure}[ht]
\includegraphics[width=0.45\textwidth]{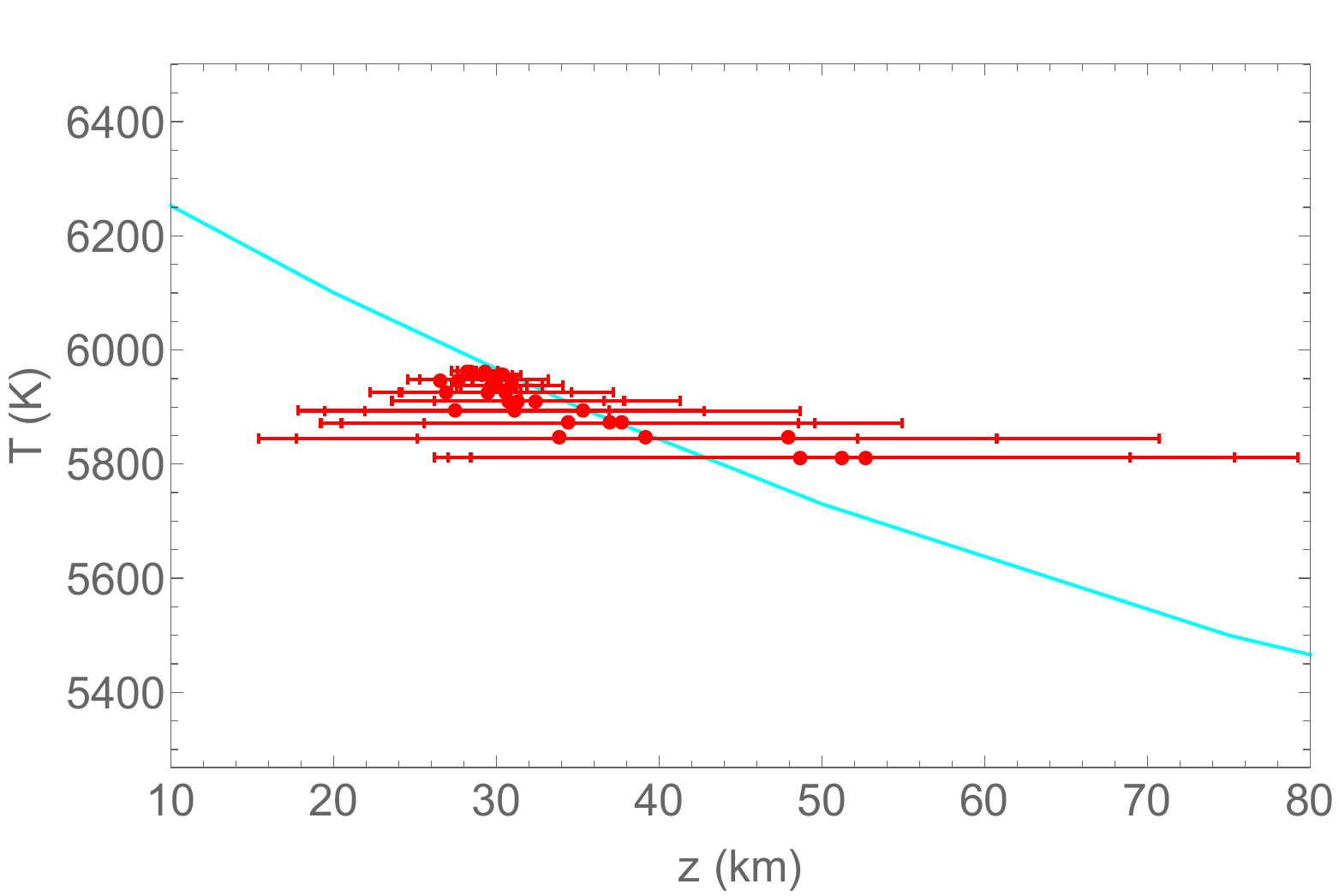}
\includegraphics[width=0.45\textwidth]{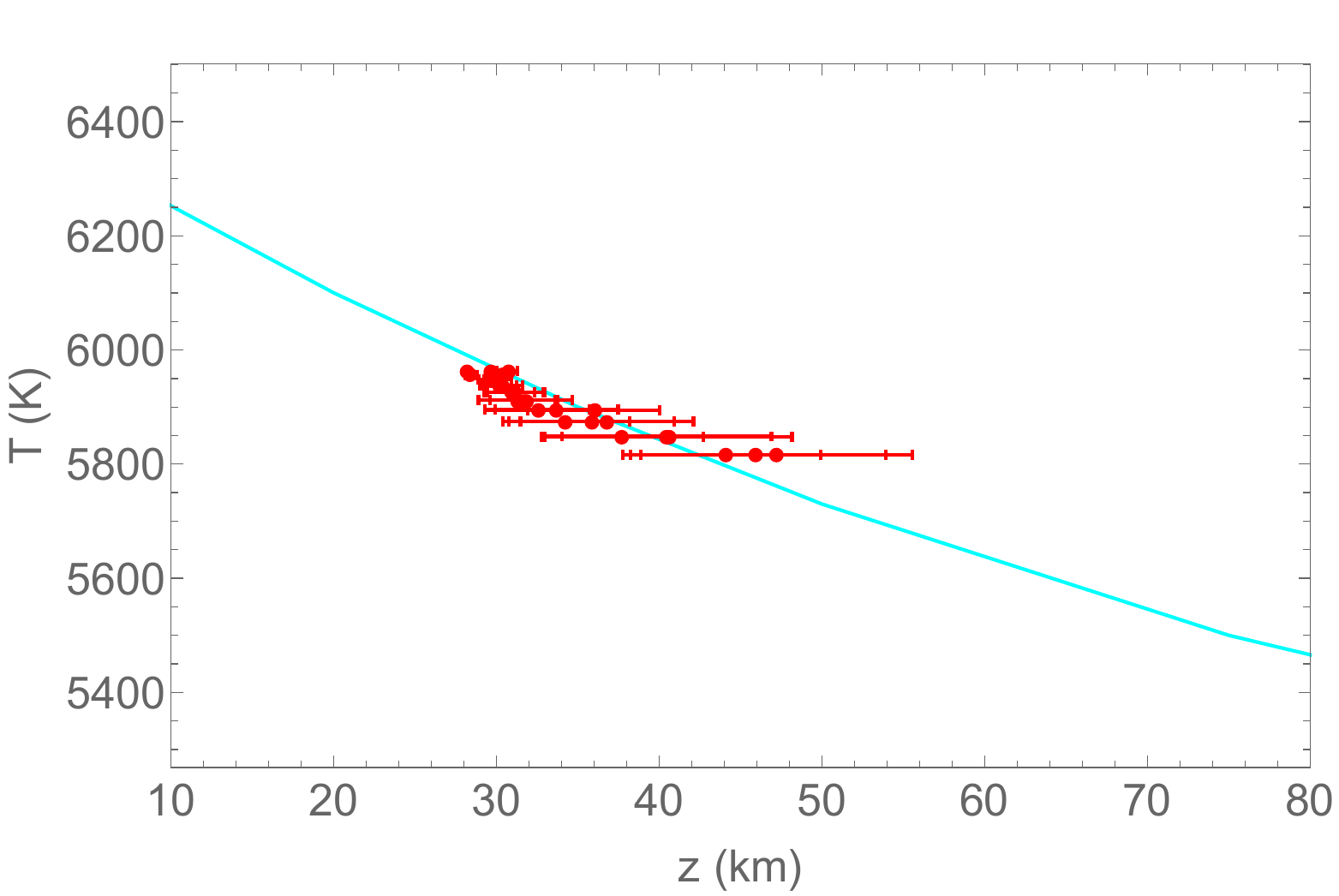}
 \caption{ Same as Fig. \ref{fig5}, but with $r_0$= 5 cm.
 }
  \label{fig6}
\end{figure}

\subsection{Measurement of the formation-depth difference}

As in \citet{Faurobert2016}, we measured  the perspective shift along the radial direction between images taken in a line wing and in the continuum
when they are observed away from disk center, that is, at heliocentric angles different from 0. 

The image obtained at a line-cord level $i$, denoted by $I_i(x,y)$, is formed higher in the photosphere than the 
continuum image $I_c(x,y)$.  Because of the perspective
effect, $I_i(x,y)$ will appear  to be shifted toward the limb in
comparison with $I_c(x,y)$ when the images are observed out of disk center. Two-dimensional images are  not needed,
and one-dimensional spectrograms may be used provided the spectrograph slit is oriented in the radial direction where the perspective shift takes place.

We  may therefore consider
one-dimensional brightness intensity variations along the slit, which is aligned with the $y$-axis in the simulations. We therefore obtained spectrograms $I_{i}(x,y)$ for
 successive values of  $x$ as if we were scanning the solar surface with a slit-spectrograph.
This allowed us to  record many
  spectrograms at \hbox{different} $x$ positions to improve the statistics
  of the solar granulation power spectrum.

Denoting for simplicity with $I_{i}(y)$  the one-dimensional cuts of
  the images, we can write
  \begin{eqnarray}\label{Claude7}
    I_{i}(y)\sim I_{c}(y - \delta)
\end{eqnarray}
\noindent and
\begin{eqnarray}\label{Claude8}
    \widehat{I}_{i}(u)\sim \widehat{I}_{c}(u)\exp(2i\pi
    u\delta).
\end{eqnarray}

 We used a series of spectrograms to estimate
the cross-spectrum $\widehat{\textit{Q}}_{ci}$ between $I_{c}(y)$
and $I_{i}(y)$:
\begin{eqnarray}\label{eq2a}
   \widehat{\textit{Q}}_{ci}(u) & = &
   <\widehat{I}_{c}(u)\widehat{I}^{*}_{i}(u)>\\ \nonumber
                             & \sim & <\mid \widehat{I}_{c}(u)\mid^{2}> e^{-2i\pi\delta
                             u} \label{eq2b},
\end{eqnarray}
where $\widehat{I}^{*}_{i}(u)$ represents the conjugate
Fourier transform of $I_{i}(y)$, and the brackets refer to the
ensemble average.

The perspective displacement introduces
 a deterministic linear phase-term $ 2 \pi \delta u$,
proportional to the spatial frequency $u$. The measurement of the phase slope gives the shift and thus the height difference 
between the surfaces where the images are formed.

We stress here that this method allows us to measure very small displacements between images, as it
is not limited by the spatial resolution of the instrument. The main limitation is the signal-to-noise
ratio of  the granulation spectrum and
the domain of validity of Eq. \ref{Claude7}, that is, of the assumption of similarity between the line wing and continuum images.

\subsection{Continuum formation depth}

The method presented above allows us to measure the formation height of  line-wing images with respect to
the continuum at 630 nm at the same heliocentric angle. To compare the temperature gradient that we measure to standard 
photospheric models, we  need to evaluate the formation height of the
630 nm continuum with respect to the conventional 500 nm continuum  formation height that defines the altitude $z=0$ in
the solar photosphere. To achieve this,  we refer to the quiet-Sun model 101 of \citet{fontenla2011},  and we assume that the temperature 
derived from the mean intensity in the continuum image follows this $(T, z)$ model. This yields the formation height of the 630 nm  continuum
at the observed heliocentric angle.
This means that the continuum point in our $(T, z)$ curves is always, by construction, on the 101 standard
model.
We stress here again that we  can only measure temperature gradients.

\section{Results} 

Figures \ref{fig5} and \ref{fig6}  show the results of the measurements  on three snapshots
of the RH simulation, with and without AO, for seeing conditions $r_0$ = 7 cm and 5 cm, respectively.  We also show for comparison 
the standard quiet-Sun model 101 of \citet{fontenla2011}. As we used RH simulations of the granulation with no magnetic field, we expect
the temperature gradient to follow the  standard quiet-Sun model. 

Images degraded by atmospheric turbulence
lead to both a bias on the temperature gradient and a loss of accuracy (the standard deviation on one measurement increases). 
When degraded images with $r_0$= 7 cm are used, the temperature gradient is underestimated, and the typical standard deviation
on the altitude is on the order of 15 km.
In the worse-case where $r_0$=5 cm, the measurement becomes hardly meaningful, with standard deviations on the order of 25 km
on the altitude.

However, the image correction by the AO system in both cases
allows us to significantly improve the quality and reliability of the measurements. 
The measured points approach the semi-empirical model very closely,
and the standard deviation decreases significantly, to reach values on the order of5 km.
The temperature gradient we measure is then consistent with the standard quiet-Sun model.

\section{Conclusion}

We have tested a method for measuring the temperature gradient in the low solar photosphere from a ground-based instrument.
The method has been applied previously  to high-resolution spectroscopic data obtained with the SOT/SP instrument on board the Hinode
satellite.
It relies on the measurement of the perspective shift between images taken at different levels in a spectral line and in the continuum when
observed out of the center of the solar disk. We emphasize that the displacements between structures
 we seek to measure are not limited by diffraction.
A well-known result of astrometry is that
 it is clearly possible to measure a displacement of an unresolved object far below
 the resolution of the telescope.  
 In our approach, the
 displacement is derived from the phase of the  cross-spectrum of  images at two different
line cords.  This information may  alternatively be derived from the cross-correlation of the
  images. This technique was first
proposed by \citet{Beckers1982} and later developed for
stellar applications \citep{Aime1984}.   

Our tests have been performed on simulated images obtained from the \textsc{Stagger}-code and in the absence of a 
magnetic field. The FeI 630.15 nm line profile was obtained from a three-dimensional LTE radiative transfer computation. Then the 
 image degradation by the atmospheric turbulence was obtained by convolution with the PSF of the atmosphere+ instrument
 without and with an AO system. The AO system and instrument characteristics were taken from the DKIST performances.
 We considered  two cases of atmospheric turbulence conditions, with Fried parameters $r_0$= 7 cm and $r_0$ = 5 cm.
 
 We stress that
  the measurement of the temperature gradient in the solar photosphere from spectroscopic observations is affected by the image degradation 
  due to atmospheric turbulence that induces  both a bias and a significant decrease in measurement accuracy. 
However, the image correction with the AO
 system allows restoring the quality of the measurement. The temperature gradient we measure on spectrograms derived from the hydrodynamical 
 simulation of the granulation is in agreement with the standard semi-empirical quiet-Sun model of the solar photosphere. This validates the measurement method
 from high-resolution spectroscopic observations, but
 corrections with an AO system are mandatory to obtain meaningful results.




\bibliography{faurob.bib}
\bibliographystyle{aa}

\end{document}